\documentclass[reprint, superscriptaddress, secnumarabic, amssymb, nobibnotes, aps, prl]{revtex4-1}

\usepackage{graphicx}
\usepackage{epstopdf}
\usepackage[T1]{fontenc}
\usepackage[latin9]{inputenc}
\usepackage{amsbsy}
\usepackage{gensymb}
\usepackage[T1]{fontenc}
\usepackage[latin9]{inputenc}
\usepackage{amsmath}
\usepackage{amssymb}
\usepackage{bbm}
\usepackage{braket}
\usepackage{xcolor}
\allowdisplaybreaks
\usepackage{graphicx}
\usepackage[colorlinks=true]{hyperref}  
\hypersetup{
    bookmarks=true,         
    unicode=false,          
    pdftoolbar=true,        
    pdfmenubar=true,        
    pdffitwindow=false,     
    pdfstartview={FitH},    
    pdftitle={Probing nodeless superconductivity in La$M$Si
    	($M$ = Ni, Pt) using muon-spin rotation and relaxation},    
    pdfauthor={a},     
    pdfsubject={},   
    pdfcreator={},   
    pdfproducer={}, 
    pdfkeywords={} {} {}, 
    pdfnewwindow=true,      
    colorlinks=true,       
    linkcolor=blue, 
    citecolor=blue,        
    filecolor=magenta,      
    urlcolor=blue           
} 
\usepackage[normalem]{ulem}

\newcommand{\figref}[1]{Fig.~\ref{#1}}

\newcommand{\tableref}[1]{Table~\ref{#1}}

\renewcommand{\approx}{\simeq}

\begin{document}
\title{\textrm{Probing nodeless superconductivity in La$M$Si
($M$ = Ni, Pt) using muon-spin rotation and relaxation}}
\author{Sajilesh.~K.~P.}
\affiliation{Department of Physics, Indian Institute of Science Education and Research Bhopal, Bhopal, 462066, India}
\author{D.~Singh}
\affiliation{ISIS Facility, STFC Rutherford Appleton Laboratory, Harwell Science and Innovation Campus, Oxfordshire, OX11 0QX, UK}
\author{A.~D.~Hillier}
\affiliation{ISIS Facility, STFC Rutherford Appleton Laboratory, Harwell Science and Innovation Campus, Oxfordshire, OX11 0QX, UK}
\author{R.~P.~Singh}
\email[]{rpsingh@iiserb.ac.in}
\affiliation{Department of Physics, Indian Institute of Science Education and Research Bhopal, Bhopal, 462066, India}

\date{\today}
\begin{abstract}
\begin{flushleft}

\end{flushleft}
Systems with strong spin-orbit coupling have been a topic of fundamental interest in condensed matter physics due to the exotic topological phases and the unconventional phenomenon they exhibit. In this particular study, we have investigated the superconductivity in the transition-metal ternary noncentrosymmetric compounds La$M$Si
($M$ = Ni, Pt) with different spin-orbit coupling strength, using muon-spin rotation and relaxation measurements. Transverse-field measurements made in the vortex state indicate that the superconductivity in both materials is fully gapped, with a conventional s-wave pairing symmetry and BCS-like magnitudes for the zero-temperature gap energies. Zero field measurements suggest a time-reversal symmetry preserved superconductivity in both the systems, though a small increase in muon depolarization is observed upon decreasing temperature. However, this has been attributed to quasi-static magnetic fluctuations.  
\end{abstract}
\maketitle

\section{INTRODUCTION}
Noncentrosymmetric (NCS) materials are unique candidates to explore exotic features like unconventional superconductivity and topologically protected surface states \cite{EBA,smidman,topo1}. These remarkable materials possess an antisymmetric spin-orbit coupling, causing the formation of two spin-dependent Fermi surfaces. This in general, can lead to the Cooper pair forming with a mixed singlet-triplet character \cite{rashba1,rashba3,vm,fujimoto,kv,rashba2}. Such a scenario can lead to zero's/multiple gaps in the energy spectrum, time-reversal symmetry breaking (TRSB) and topologically protected nontrivial surface states \cite{Bauer2004,CPS,LPB2,edg1,edg2}. A breakthrough discovery occurred when the line nodes in heavy fermion compound CePt$ _{3} $Si were found,  followed by triplet pairing and nodes in weakly correlated Li$ _{2} $(Pd,Pt)$ _{3} $B \cite{Bauer2004,LPB2}. This has shown immense potential of NCS materials  to host unconventional superconductivity. Li$ _{2} $(Pd,Pt)$ _{3} $B is one of the most acclaimed compound where the antisymmetric spin orbit coupling (ASOC) effects have been directly observed. Li$ _{2} $Pt$ _{3} $B has shown the presence of triplet and line node, while Li$ _{2} $Pd$ _{3} $B with same structure has shown conventional s-wave behaviour  \cite{LPB2,LPB3,LPB4}. The unusual properties of Li$ _{2} $Pt$ _{3} $B is attributed to the increase in ASOC which is proportional to Z$ ^{4} $. Besides this, line nodes in superconducting gap are discovered for CeIrSi$ _{3} $ \cite{CeIS}, K$ _{2} $Cr$ _{3} $As$ _{3} $ \cite{KCA1,KCA2}, while multiple nodeless gap were shown by LaNiC$ _{2} $ \cite{LNC1}, (La,Y)$ _{2} $C$ _{3} $ \cite{LC}.In addition,  these materials are expected to possess topologically protected flat zero-energy bands of surface states, which can be termed as a long-sought Majorana fermion \cite{MZhasan,topo2,Hsoc2,Hsoc3,TOS}.   
\\ 

A few transition metal superconductors including La$ _{7} $Ir$ _{3} $ \cite{LI}, Re$ _{6} $X (X = Zr, Hf, Ti) \cite{RZ,RH,RT} with strong ASOC has shown a spontaneous field upon entering the superconducting state and hence TRSB. Meanwhile, LaNiC$ _{2} $ with noncentrosymmetric structure and low ASOC has also shown TRSB \cite{LNC2, LNC3}. Moreover, in a similar case, LaNiGa$ _{2} $, with centrosymmetric structure and low ASOC, has shown spontaneous field in the superconducting state, questioning the role played by ASOC \cite{LNG}. Also, many materials with considerably large ASOC have failed to show any non-trivial superconductivity \cite{RW, MAC, CIS,BPS,LPG}.\\

 The recent discovery of TRSB in pure Re metal has further increased the curiosity in this field, raising more open questions regarding the emergence of TRSB \cite{Re}. The strength of the TRSB signal in Re based binary systems (Re$ _{6} $X) has remained unaltered, irrespective of the transition element used. Though the microscopic origin of TRSB in these compounds is still unknown,  the local electronic structure of Re might be playing a crucial role.  Hence it is important to search for more NCS materials with different ASOC strength to elucidate the effects of ASOC on the superconducting ground state.\\
 \\
Since significant spin-orbit coupling is a proposed necessary criterion to exhibit exotic properties, we turned our focus onto La based NCS systems, LaNiSi and LaPtSi. Both of them crystallize in LaPtSi type structure, while the spin-orbit coupling has different values as Pt is a heavier element compared to Ni \cite{LPS1,LNS,LPS2}. Pt being a $ d $-block element with the third largest atomic number is expected to induce a stronger ASOC. A recent theoretical study on similar structure compound, Th$ T $Si ($ T $= Co, Ir, Ni, and Pt) have shown that ASOC has caused splitting of the Fermi surface into two non-degenerate sub-bands with different helicity \cite{ThTSi}. It was also noticed that this effect is stronger for the case of Ir and Pt, which are having heavier mass. Hence a microscopic investigation on La$ M $Si can explicate the effect of similar Fermi surface splitting on the superconducting ground state. Furthermore, the contribution to the electronic density of states at Fermi energy, including the spin-orbit coupling, is dominated by Pt-$ d $ band \cite{LPS3}. Hence, a comparative study with lighter Ni atom in place of Pt gives an opportunity to unravel the effects of ASOC and density of states on the superconducting ground state. Here, we have used muon spin rotation/relaxation measurement ($ \mu $SR) to investigate the superconducting ground state. Zero field $ \mu $SR is exceptionally sensitive to intrinsic local magnetization arises at superconducting phase transition in case of unconventional pairing mechanism. Besides, the transverse field $ \mu $SR is an excellent tool to probe the superconducting gap structure. It can accurately probe the penetration depth in superconductors, and measuring the temperature dependence provides details of the gap structure. The technique has already used in unraveling the unconventional nature of many superconductors. It has been widely used in materials including heavy fermion superconductors \cite{UPt,PrOsSb}, Fe based superconductors \cite{FeSe}, other alloy based superconductors \cite{LI,RZ,RH,RT,Re} giving path breaking results.

\section{EXPERIMENTAL METHODS}

Polycrystalline samples of LaNiSi and LaPtSi were prepared by arc melting stoichiometric amounts of the constituent elements on a water-cooled copper hearth under the argon gas atmosphere. The samples were flipped and remelted several times to ensure the homogeneity of the ingot. There was no measurable weight loss during the melting. All samples were wrapped in Ta foil, sealed in quartz ampoules under vacuum, and annealed at 800 \textdegree{}C for one week to remove any thermal strain. The sample characterization was done using the x-ray powder diffraction (XRD) on a PANalytical diffractometer using the Cu K$_{\alpha}$ radiation ($\lambda$ = 1.54056 $\text{\AA}$). Magnetic susceptibility measurements were performed on a Magnetic Property Measurement System (MPMS) Superconducting Quantum Interference Device (SQUID) magnetometer (Quantum Design). Heat capacity measurements were performed using Quantum Design Physical Property Measurement System (PPMS). The muon-spin relaxation/rotation ($\mu$SR) measurements were carried out using the MUSR spectrometer at the ISIS Neutron and Muon facility in STFC Rutherford Appleton Laboratory, United Kingdom. The powdered samples of La$M$Si ($M$ = Ni, Pt) were mounted on a high-purity-silver plate using diluted GE varnish. For LaNiSi, the measurements were performed in the temperature range 0.1 K - 2.0 K, whereas, for LaPtSi, the measurements were made between 0.1 K and 4.0 K. The $\mu$SR measurements were performed under the longitudinal and transverse-field geometry. During measurement, spin-polarized muons were implanted into the sample. The implanted muons precess according to the local magnetic field distribution and emit positrons during decay after a lifetime of 2.2 $ \mu $s. The distribution of positrons gives vital information regarding the nature of internal field distribution. In zero field configuration, the stray fields at the sample position due to neighboring instruments and the Earth's magnetic field is canceled to within $\sim$ 1.0 $\mu$T using three sets of orthogonal coils. In the transverse configuration, a field was applied perpendicular to the direction of the muon spin (which is opposite to muons linear momentum), and the detectors were grouped into two orthogonal pairs.  A full description of the $\mu$SR technique may be found in Ref. \cite{SLL}. 

\begin{figure}
	\includegraphics[width=1.0\columnwidth]{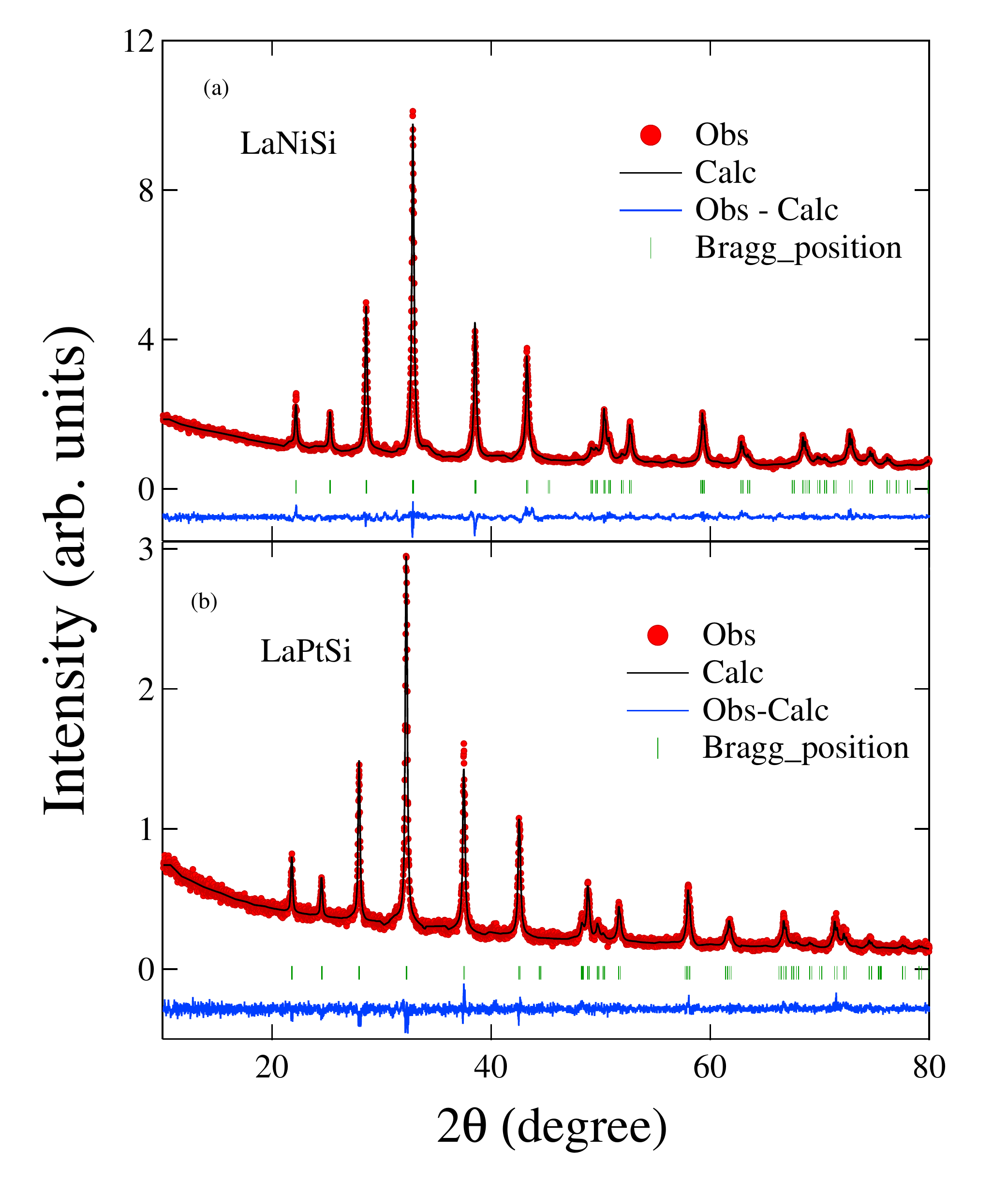}
	\caption{\label{XRD} X-ray diffraction pattern collected at ambient conditions for (a) LaNiSi and (b) LaPtSi refined with noncentrosymmetric $ \alpha $-ThSi$ _{2} $ type structure. }
\end{figure}

\section{RESULTS AND DISCUSSION}

\begin{figure}
\includegraphics[width=1.0\columnwidth]{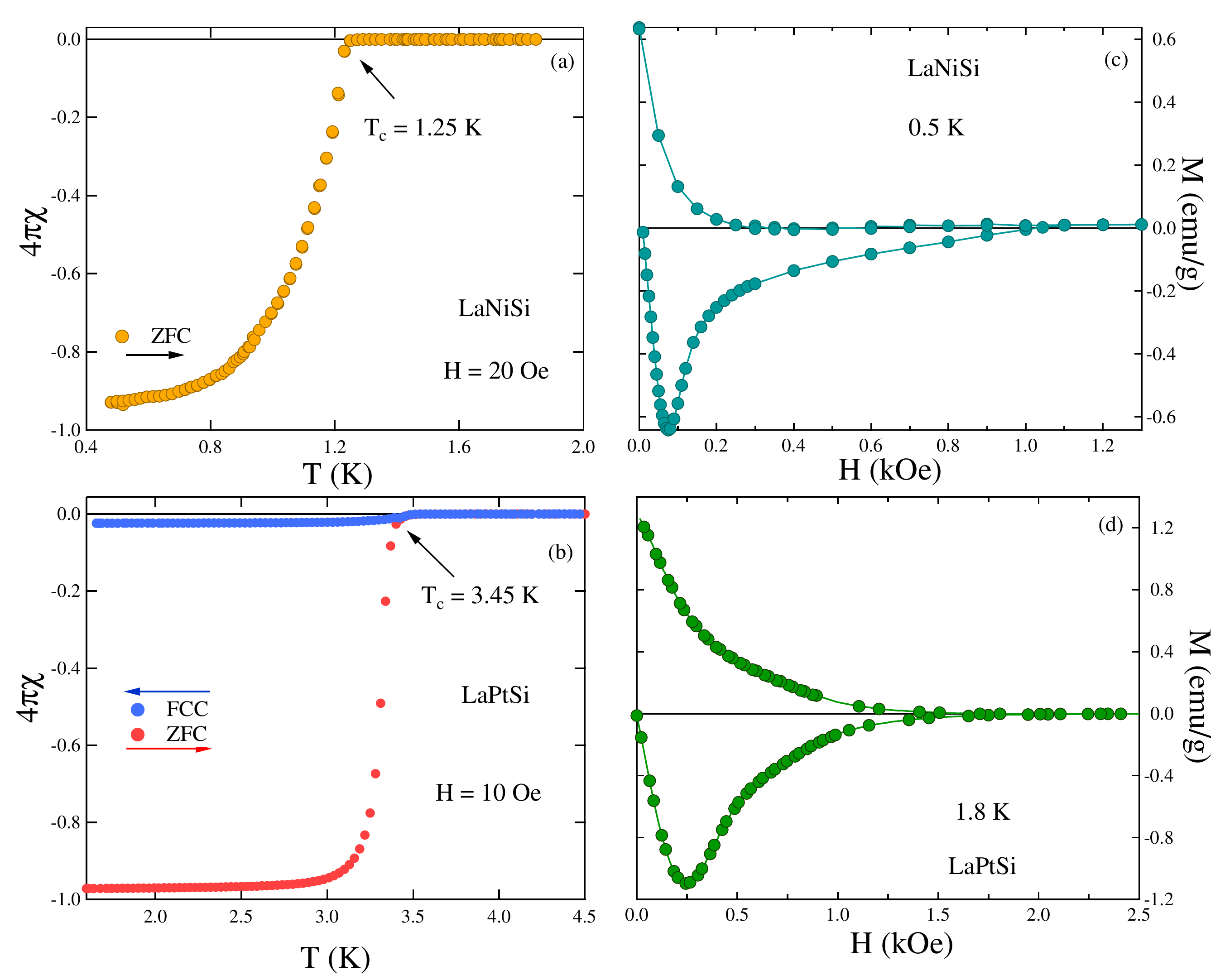}
\caption{\label{MT}(a) and (b). Magnetization data collected at ZFC-FC mode showing the superconducting transition at 1.25 K and 3.45 K respectively for LaNiSi and LaPtSi. (c) and (d) M-H curve taken at superconducting regime showing a type-II behavior by both compounds.}
\end{figure}

Powder x-ray diffraction data were collected for both the samples. Rietveld refinement of the data confirmed that both samples had crystallized into the tetragonal, noncentrosymmetric structure with space group $I 4_{1}md$ (109) (\figref{XRD}). The lattice parameters of LaNiSi (a=b= 4.1800(3) \text{\AA}, c = 14.0780(8) \text{\AA}) and LaPtSi (a=b= 4.2466(8) \text{\AA}, c = 14.5213(4) \text{\AA} ) obtained in this work are in good agreement with data reported previously in Refs. \cite{LNS,LPS2}.\\
The samples were characterized using the dc susceptibility measurements in zero-field-cooling and field-cooling modes under an applied magnetic field. Appearance of a strong diamagnetic signal at T$_{c}$ = 1.25 $\pm$ 0.02 K in LaNiSi and T$_{c}$ = 3.45 $\pm$ 0.04 K in LaPtSi confirms the bulk superconducting nature (\figref{MT} (a) and (b)). The Meissner volume fraction 4$\pi\chi$ for both samples are less than 100\%  due to uncorrected geometrical shape factor. Magnetization measurements exhibit no other magnetic anomalies  that may be due to impurities in the sample. 

\figref{MT}(c) and (d) shows the magnetization data collected below the superconducting transition temperature for both the samples. Magnetization data shows a linear behavior at low field, after which the sample enters a vortex state. This clearly depicts the type-II nature of the sample.

\begin{figure}
	\includegraphics[width=1.0\columnwidth]{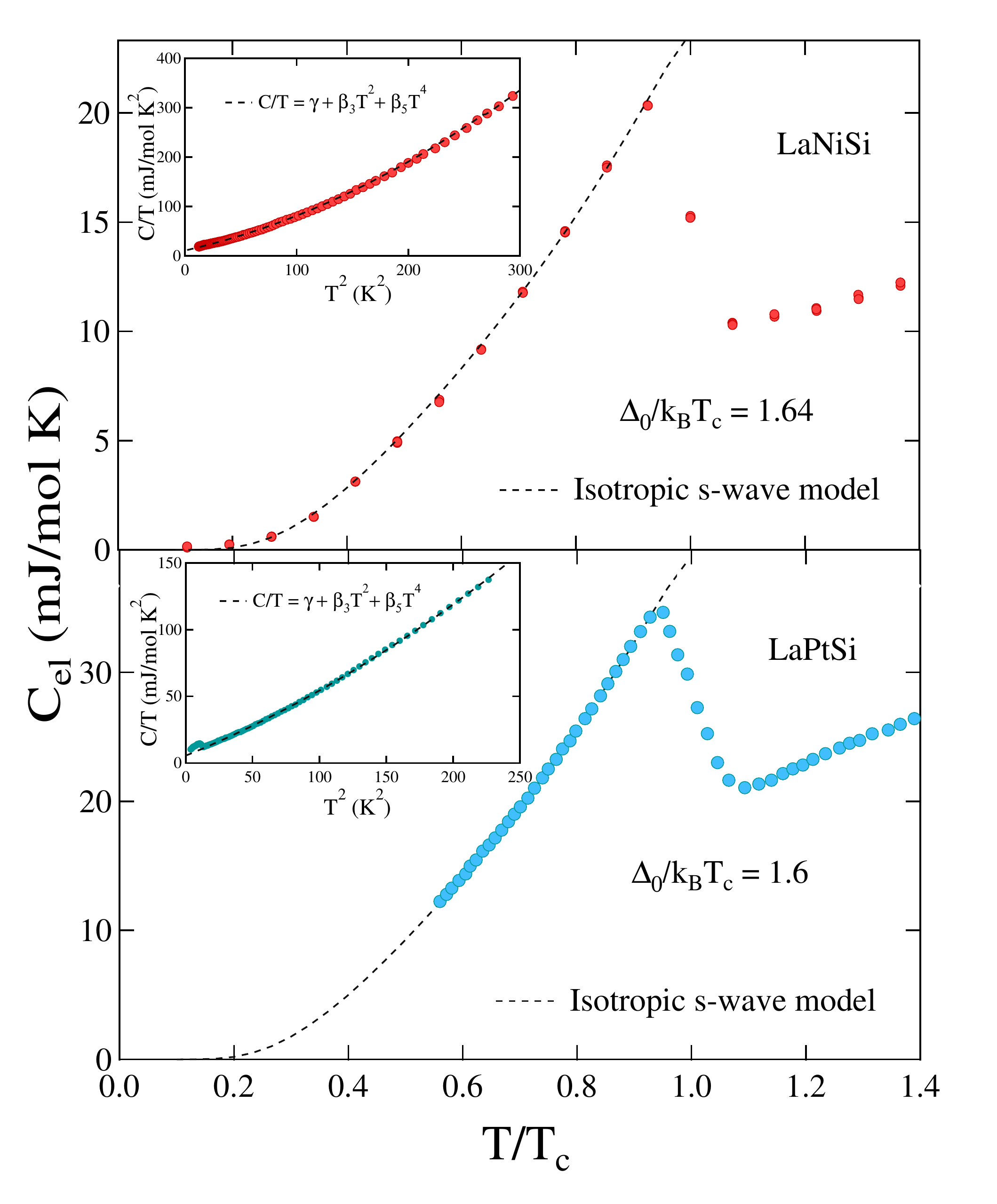}
	\caption{\label{SP} Temperature dependent electronic specific heat data for both LaNiSi and LaPtSi taken at 0 T. The superconducting region can be well traced by isotropic BCS s-wave model, giving the normalized specific heat jump as 1.64 and 1.6 respectivly for LaNiSi and LaPtSi. The insets shows the total specific heat data plotted as C/T Vs T$ ^{2} $.}
\end{figure}

Further investigation on the superconducting nature was done by temperature dependent specific heat analysis. A superconducting anomaly for both the samples were observed at T$ _{c} $ = 1.11 K and 3.4 K (\figref{SP}). The normal state specific heat for the samples above T$ _{c} $ can be described by C = $ \gamma $T + $ \beta_{ 3 } $T$ ^{3} $ + $ \beta_{5 }$ T$ ^{5} $. This gave the fitting parameters as $ \gamma_{n}$ = 9.12 $\pm$ 0.07  mJ/molK$^{2}$, $\beta_{3}$ = 0.487 $\pm$ 0.001  mJ/mol K$^{4}$ and $ \beta_{5} $ = (1.98 $ \pm $ 0.01)$ \times $ 10$ ^{-4} $mJ/mol K$^{6}$ for LaNiSi while for LaPtSi it is $ \gamma_{n}$ = 4.72 $\pm$ 0.22  mJ/mol K$^{2}$, $\beta_{3}$ = 0.423 $\pm$ 0.004  mJ/mol K$^{4}$ and $\beta_{5}$ = (7.34 $ \pm $ 0.01)$ \times $ 10$ ^{-4} $mJ/mol K$^{6}$. Several parameters characterizing the materials can be deduced using these values and shown in table \ref{elec}. The electronic specific heat in superconducting region is well explained by an isotropic s-wave model, giving the normalized superconducting gap, $ \Delta_{0} $/k$ _{B} $T$ _{c} $ = 1.64 $ \pm $ 0.04 and 1.61 $ \pm $ 0.05 respectively for LaNiSi and LaPtSi.

\begin{figure}
	\includegraphics[width=1.0\columnwidth]{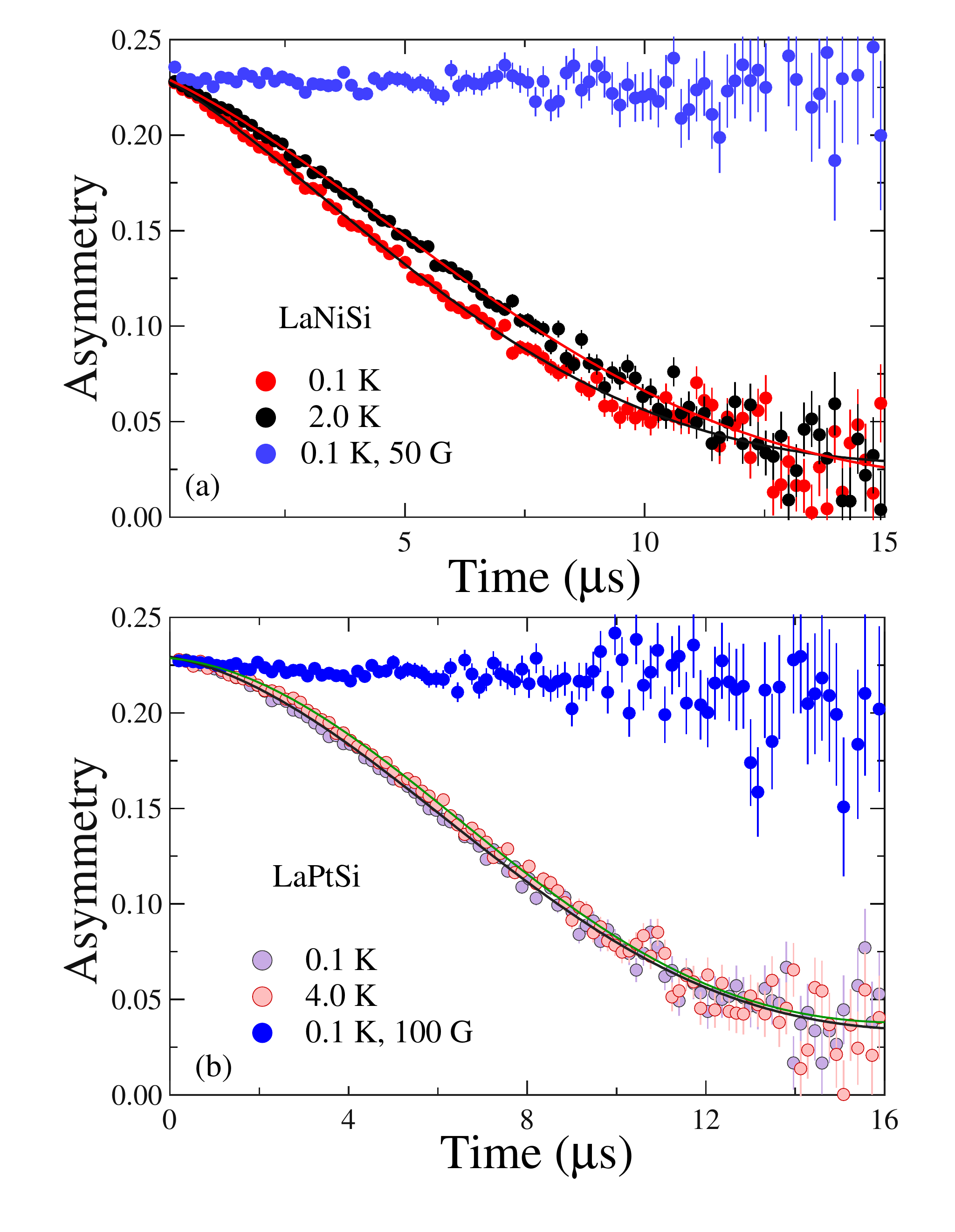}
	\caption{\label{ZF} Time evolution of the spin polarization of muons implanted
		under zero-field conditions in (a) LaNiSi and (b) LaPtSi at
		temperatures above and below T$_{c}$. The solid lines are the results of fitting the data to Eq. (2). Blue markers show the muon depolarization at a small longitudinal applied field.}
\end{figure}
\begin{figure}
	\includegraphics[width=1.0\columnwidth]{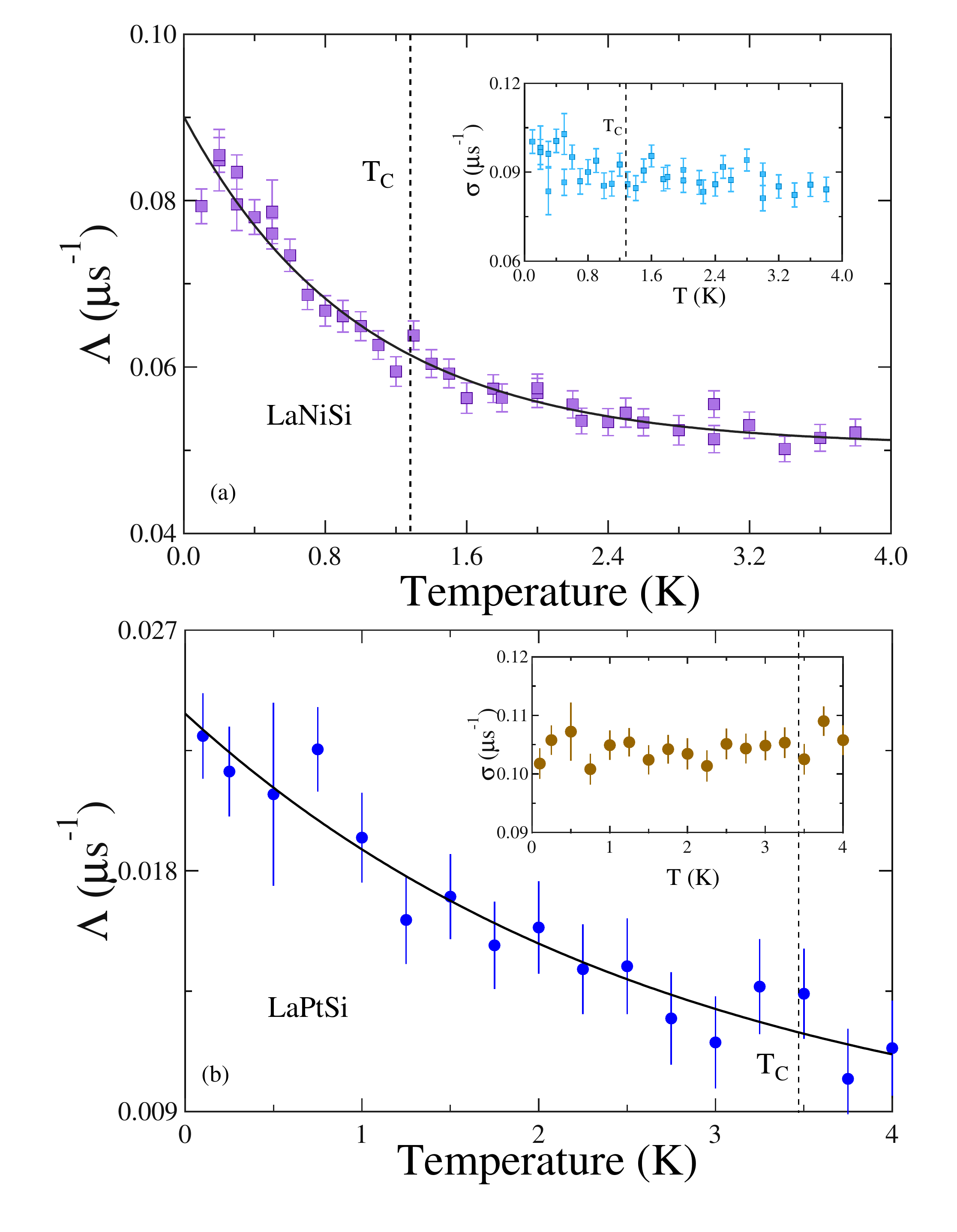}
	\caption{\label{lamda} Temperature dependence of the electronic relaxation rate
		in (a) LaNiSi and (b) LaPtSi, collected in ZF. The solid lines are guides to the eye, indicating the exponential decay of $\Lambda$ in ZF as T is increased. The insets show the constant behavior of nuclear relaxation rate $ \sigma $ across the transition temperature.}
\end{figure}

A systematic ZF-$\mu$SR measurements can be used to detect any spontaneous magnetization below the superconducting transition.   We performed the ZF-$\mu$SR relaxation experiments on both La$M$Si ($M$ = Ni, Pt) samples, and \figref{ZF} shows the ZF-$\mu$SR spectra for both samples at selected temperatures above and below T$_{c}$. Below T$_{c}$, there is a clear change in the relaxation behavior in both compounds. The relaxation became faster with decreasing temperature down to the lowest temperature, although the difference is much subtler in LaPtSi. Notably, there is no sign of an oscillatory component which would otherwise indicate coherent field associated with magnetic ordering. In the absence of atomic moments, the relaxation is due to the presence of static, randomly oriented local fields associated with the nuclear moments. The ZF-$\mu$SR data can be well described by a damped Gaussian Kubo-Toyabe (KT) function:
\begin{equation}
G(t)= A_{1}\mathrm{exp}(-\Lambda t)G_{\mathrm{KT}}(t)+A_{\mathrm{BG}} ,
\label{eqn2}
\end{equation}
where $A_{1}$ is the sample asymmetry, $\Lambda$ is the additional relaxation rate, and $A_{\mathrm{BG}}$ is the background asymmetry. The G$_{\mathrm{KT}}$(t) function is the Gaussian Kubo-Toyabe function given by \cite{KT1}:
\begin{eqnarray}
G_{\mathrm{KT}}(t) &=&\frac{1}{3}+\frac{2}{3}(1-\sigma^{2}_{\mathrm{ZF}}t^{2})\mathrm{exp}\left(\frac{-\sigma^{2}_{\mathrm{ZF}}t^{2}}{2}\right),
\label{eqn3}
\end{eqnarray}
where $\sigma_{\mathrm{ZF}}/\gamma_{\mu}$ is the local field distribution width, $\gamma_{\mu}$ = 135.53 MHz/T being the muon gyromagnetic ratio. The parameters $A_{1}$ and $A_{\mathrm{BG}}$ extracted by fitting the ZF-$ \mu $SR spectra using Eq. \eqref{eqn2} are found to be temperature independent for both the samples. The temperature dependence of the fit parameters $\sigma_{\mathrm{ZF}}$ and $\Lambda$ for LaNiSi and LaPtSi are displayed in \figref{lamda} (a) and (b). Again, the nuclear term $\sigma_{\mathrm{ZF}}$ is found to be approximately temperature independent in both compounds [see \figref{lamda} insets]. In contrast, the additional relaxation rate, $\Lambda$, seen to be increasing gradually with decreasing temperature [shown in \figref{lamda}]. There is no distinct anomaly at T$_{c}$. Therefore, the observed behavior, we believe, could not be associated with the superconducting nature of the samples. The exponential character of $\Lambda$(T) in both materials reveals the existence of fast electronic fluctuations measurable within the $\mu$SR time window.
A decrease in fluctuation frequency of electronic moments as temperature decreases may cause $ \Lambda $ to increase. Similar behavior is observed in number of superconductors \cite{RRuB,ReBe,HVG}. However, the exact nature and source of this behavior is still unknown and therefore require further investigation. The nature of ZF relaxation can be further explored by the application of the longitudinal field. In both materials, a small longitudinal field is sufficient to completely decouple the static fields, with the overall depolarization being minimized. This implies that the fluctuations responsible for this relaxation channel are in fact static or quasi-static with respect to the muon lifetime and the magnitude of the fluctuations is $\le$ 100 Oe.\\

\begin{figure*}
	\includegraphics[width=2.0\columnwidth]{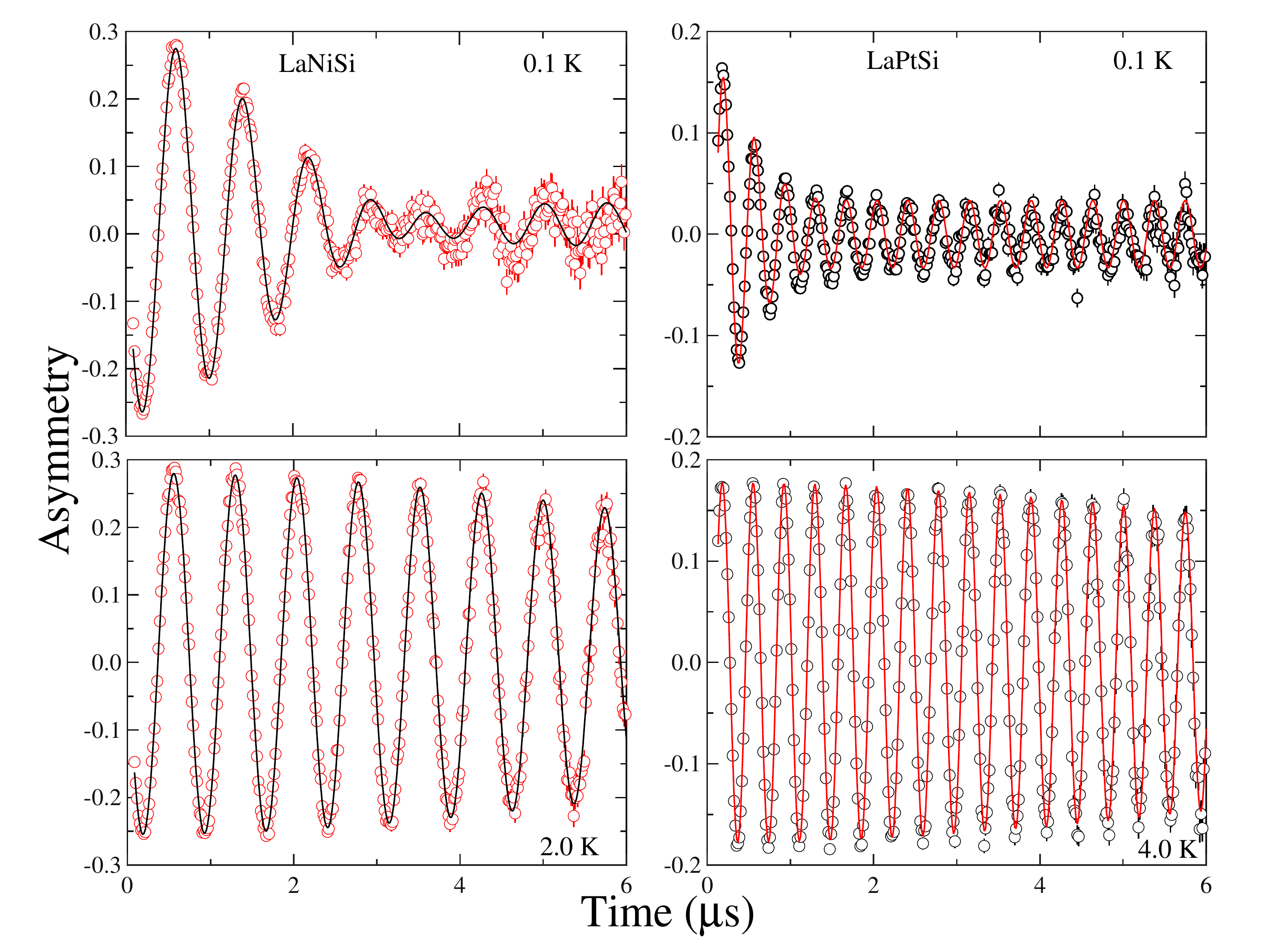}
	\caption{\label{TF} Transverse-field muon time spectra collected (a) in a magnetic field H = 100 Oe at 0.1 K and 2.0 K for the LaNiSi and (b) in a magnetic field H = 200 Oe at 0.1 and 4.0 K for the LaPtSi.}
\end{figure*}
\begin{figure} 
\includegraphics[width=9. cm, height=12. cm, origin=b]{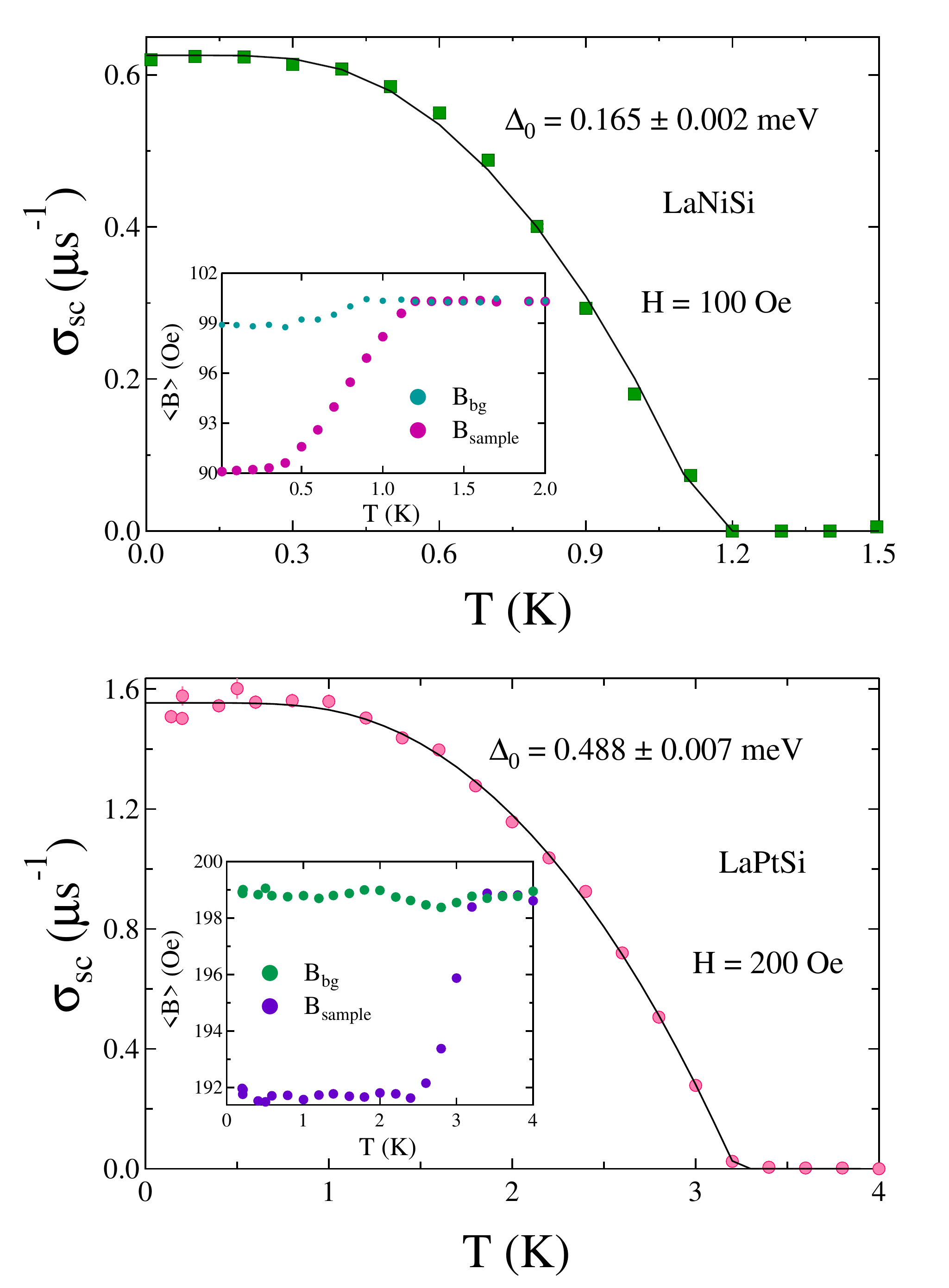}
\caption{Temperature dependence of $ \sigma_{sc} $ measured at an applied field. The solid line is the dirty limit isotropic s-wave fit for the data. The inset shows the internal field variation for the samples.}
\label{fig5}
\end{figure}

TF-$\mu$SR is an excellent tool to explore the gap structure of superconducting materials. We have performed a TF-$\mu$SR experiment down to a temperature of 0.1 K in order to probe the flux line lattice and therefore determine the symmetry of the superconducting gap. \figref{TF} shows the $\mu$SR precession signals below and above T$_{c}$ for both the LaNiSi and LaPtSi compounds. The data were collected in an applied field of H = 100 Oe for LaNiSi and H = 200 Oe for LaPtSi. The field was applied above T$_{c}$ before cooling down to 0.1 K in order to ensure that the samples are in the mixed state. Figures \ref{TF}(b) and \ref{TF}(d) show the spectra above T$_{c}$ for both the samples where the spectra oscillate with a frequency that corresponds to the Larmor precession, damped with a weak Gaussian relaxation due to the nuclear dipole field. Below T$_{c}$, the signal decays with time due to the inhomogeneous field distribution from the flux line lattice [shown in \figref{TF}(a) and \figref{TF}(c)]. To quantitatively analyze the experimental data, the following oscillatory decaying Gaussian function was employed:
\begin{eqnarray}
G_{\mathrm{TF}}(t) &=& A_{0}\mathrm{exp}\left(\frac{-\sigma^{2}t^{2}}{2}\right)\mathrm{cos}(\omega_{1}t+\phi)\nonumber\\&+&A_{1}\mathrm{cos}(\omega_{2}t+\phi) .
\label{eqn3}
\end{eqnarray}
Here $A_{0}$ and $A_{1}$ are the initial asymmetries of the sample and background signals, $\omega_{1}$ and $\omega_{2}$ are the precession frequencies of muons from the sample and silver holder, respectively, $\phi$ is the phase offset of the initial muon spin polarization with respect to positron detector. and $\sigma$ is the depolarization rate. The inset of \figref{fig5} shows the temperature dependence of internal magnetic field, calculated from the muon precession frequency. The flux expulsion at T$_{c}$ is evident from the reduction of the average field $<B>$ inside the superconductor, and the corresponding background field $B_{bg}$ is approximately constant over the temperature range. The muon depolarization rate $\sigma$ extracted from Eq. \ref{eqn3} is comprised of the following terms: $\sigma^{2}$ = $\sigma_{\mathrm{sc}}^{2}+\sigma_{\mathrm{N}}^{2}$, where $\sigma_{\mathrm{sc}}$ is the depolarization arising due to the field variation across the flux line lattice and $\sigma_{\mathrm{N}}$ is the contribution due to nuclear dipolar moments. The superconducting contribution to depolarization  $\sigma_{\mathrm{sc}}$ is calculated by subtracting $\sigma_{\mathrm{N}}$  from total  $\sigma$.\\

The temperature dependence of $\sigma$ is seen nearly constant below $\approx$ T$_{c}$/3  for both compounds. This possibly suggests the absence of nodes in the superconducting energy gap at the Fermi surface. The solid line in \figref{fig5} represents the temperature dependence of the muon depolarization rate $\sigma(T)$ within the local London approximation for a s-wave BCS superconductor in the dirty limit:
\begin{equation}
\frac{\sigma_{FLL}^{-2}(T)}{\sigma_{FLL}^{-2}(0)} = \frac{\Delta(T)}{\Delta(0)}\mathrm{tanh}\left[\frac{\Delta(T)}{2k_{B}T}\right] ,
\label{eqn5}
\end{equation}
where $\Delta(T)/\Delta(0) = \tanh\{1.82(1.018({T_{c}/T}-1))^{0.51}\}$ is the BCS approximation for the temperature dependence of the energy gap and $\Delta(0)$ is the gap magnitude at zero temperature. While in the clean limit, the expression is given  by
\begin{equation}
\frac{\sigma_{FLL}^{-2}(T)}{\sigma_{FLL}^{-2}(0)} = 1+2\int_{\Delta(T)}^{\infty}\left(\frac{\delta f}{\delta E}\right)\frac{E dE}{\sqrt{E^{2}-\Delta^{2}(T)}}  ,
\label{eqn5}
\end{equation}

Here, $ f $ = [1+exp(E/k$_{B}T)]^{-1} $ is the Fermi function and $ \Delta (T) =  \Delta_{0}\delta (T/T _{c} )$. $\delta  (T/T _{c} )$ = $\tanh\{1.82(1.018({T_{c}/T}-1))^{0.51}\}$ is the temperature dependence of the energy gap.
We have obtained a good fit for the data using the the dirty limit model giving values of the superconducting gap as $ \Delta_{0} $ = 0.197 $ \pm $ 0.003 meV and 0.488 $ \pm $ 0.007 meV for LaNiSi and LaPtSi respectively. This gives the normalised value of superconducting gap  $ \frac{\Delta(0)}{k_{B} T_{c}} $ as 1.63 and 1.74 respectively for LaNiSi and LaPtSi, showing a moderately coupled nature of samples consistent with previous reports \cite{LNS,LPS2}.  

In a superconductor with ideal Ginzburg-Landau vortex lattice , Brandt has explained the relation between the magnetic penetration length $ \lambda $ and muon depolarization rate $ \sigma_{sc} $ \cite{brandt1,brandt2}. According to this, for a superconductor with h = $ H/H_{c2} $ $ \leq $ 0.25, 
\begin{equation}
\sigma_{\mathrm{FLL}} [\mu s^{-1}] = 4.854\times 10^{4}(1-h)[1+1.21(1-\sqrt{h})^{3}]\lambda^{-2} [nm^{-2}]
\label{bran}
\end{equation}

Substituting the value of $ \sigma_{sc} $ gives $ \lambda $ = 352 $ \pm $ 19 nm and 226 $ \pm $ 11 nm for LaNiSi and LaPtSi respectively. Using this, we estimated the superconducting carrier density n$ _{s} $, by $ n_{s}(0) = \frac{m^{*}}{\mu_{0}e^{2}\lambda^{2}} $, where $ m^{*} = (1+\lambda_{e-ph})m_{e} $. We have used $ \theta_{D} $ obtained from specific heat measurement to calculate $ \lambda_{e-ph} $ = 0.48 $ \pm $ 0.02 for LaNiSi and  $ \lambda_{e-ph} $ = 0.61 $ \pm $ 0.02 for LaPtSi \cite{NbOs}. Substituting this has given the superconducting carrier density as (3.37 $ \pm $ 0.36)$ \times $10$ ^{26} $/m$ ^{3} $ and (8.85 $ \pm $ 0.86 ) $ \times $10$ ^{26} $/m$ ^{3} $ for LaNiSi and LaPtSi respectively. This can be used to calculate the Fermi temperature for the materials using the relation,

\begin{equation}
 k_{B}T_{F} = \frac{\hbar^{2}}{2}(3\pi^{2})^{2/3}\frac{n_{s}^{2/3}}{m^{*}}, 
\label{eqn13:tf}
\end{equation} 

\begin{figure}
	\includegraphics[width=1.0\columnwidth]{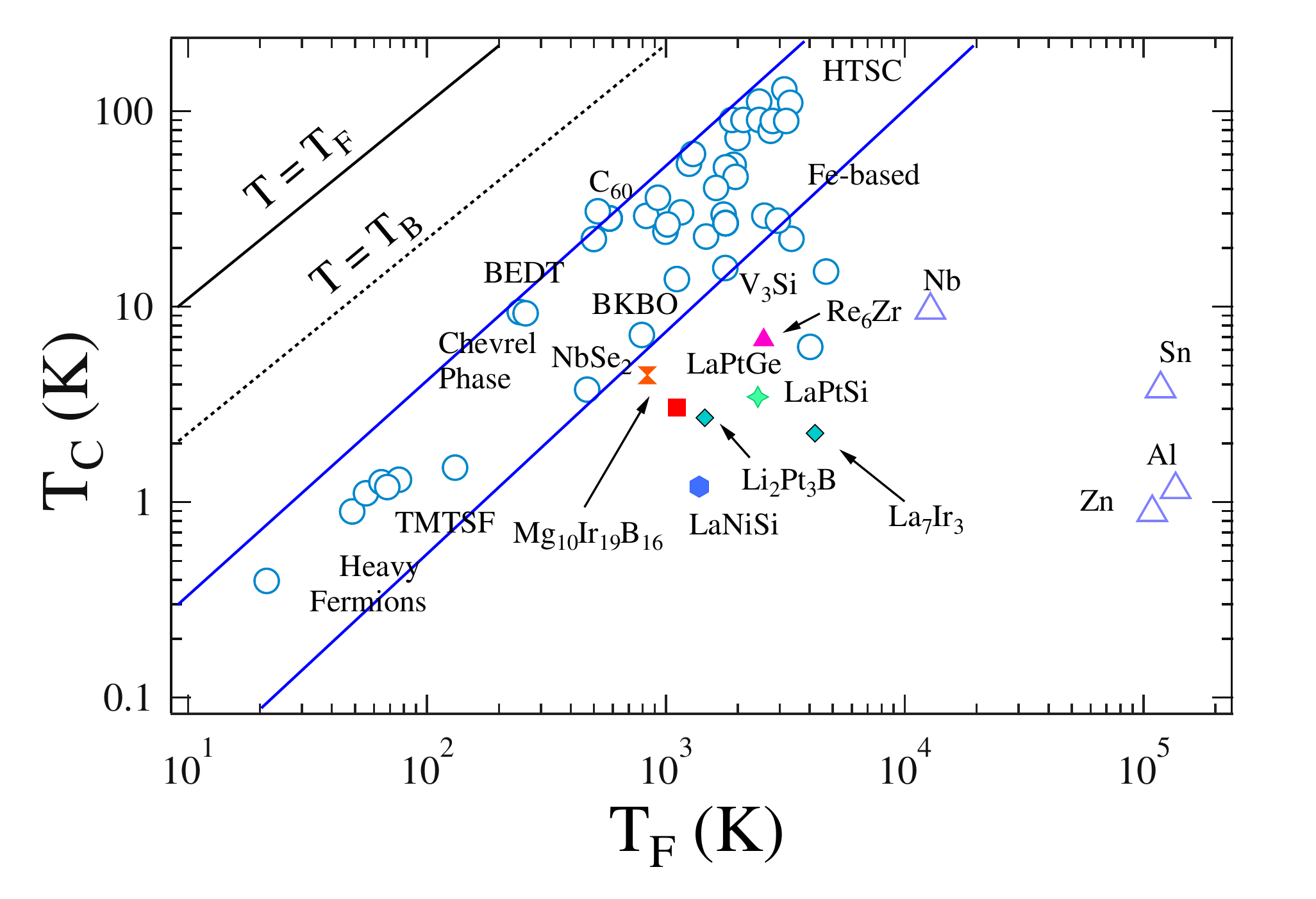}
	\caption{\label{Uemura}Uemura plot showing  T$_{c} $ Vs. the effective Fermi temperature T$ _{F} $. The blue band represents different families of superconductors with unconventional properties. The positions of LaNiSi and LaPtSi indicate a conventional nature of these materials.}
\end{figure}

The obtained values for T$_{F} $ are 1380 $ \pm $ 98 K and 2430 $ \pm $ 157 K respectively. The values are close to those reported elsewhere for transition metal alloys \cite{NbOs}. This can be further used to classify the superconductors as done by Uemura $ et$ $al. $ \cite{umera1,umera2,umera3,umera4}. According to the Uemura classification scheme, high-temperature superconductors, heavy fermionic superconductors, Fe-based superconductors, and other exotic superconductors falls in the range 0.01$ \leq $$ \frac{T_{c}}{T_{F}} $$ \leq $0.1.  For conventional BCS superconductors, $ \frac{T_{c}}{T_{F}} \leq$ 0.001. The value of $ \frac{T_{c}}{T_{F}} $ = 0.0008 and 0.0014 for LaNiSi and LaPtSi places both the compounds away from the unconventional band of superconductors, as shown in  \figref{Uemura}, but close to other superconductors that may be considered as exotic, such as the nickelborocarbides \cite{umera4}. Calculated superconducting parameters are tabulated in \tableref{elec}.

\begin{table}[h!]
\caption{Normal and superconducting properties of La(Ni,Pt)Si}
\label{elec}
\begin{center}
\begin{tabular*}{1.0\columnwidth}{l@{\extracolsep{\fill}}llll}\hline\hline
Parameter& Unit& LaNiSi & LaPtSi\\
\hline
\\[0.5ex]                                        
$T_{c}$&K& 1.25 $ \pm $ 0.02 & 3.45 $ \pm $ 0.04 \\ 
$ \theta_{D} $&K&230 $ \pm $3&239 $ \pm $2\\
$ \lambda_{e-ph} $ & &0.48 $ \pm $ 0.02&0.61 $ \pm $ 0.02\\
D$ _{c} $($E _{f} $)&$ \frac{states}{eV.f.u} $& 3.87 $ \pm $ 0.12 &  1.99 $ \pm $ 0.06\\        
$\Delta(0)/k_{B}T_{c}$& & 1.63 $ \pm $ 0.04 & 1.74 $ \pm $ 0.05\\
$m^{*}/m_{e}$& & 1.48 $ \pm $ 0.06 & 1.61 $ \pm $ 0.13\\             
n& 10$ ^{26} $m$ ^{-3} $& 3.37  $ \pm $ 0.36 & 8.85 $ \pm $ 0.86\\                     
$\lambda_{L}$& nm & 352 $ \pm $ 19& 226 $ \pm $ 11\\
$T_{F}$& K & 1380 $ \pm $ 90 & 2430 $ \pm $ 157\\
\\[0.5ex]
\hline\hline
\end{tabular*}
\par\medskip\footnotesize
\end{center}
\end{table}

\section{Conclusion}
We have probed the superconducting properties of ternary equiatomic silicides La$ M $($ M $=Ni, Pt)Si by magnetization, specific heat and muon spin rotation and relaxation measurements. The specific heat measurements have indicated an s-wave nature for both the samples. A systematic TF-$ \mu $SR study at an applied field reveals the T$ _{c} $ for the samples as 1.2 K and 3.45 K respectively for LaNiSi and LaPtSi with both showing a type-II nature. A temperature independent nature of muon depolarization rate at low temperatures ruled out the presence of any anisotropic or nodal nature of the superconducting gap. The well-fitted data using the isotropic s-wave model has revealed a moderately coupled nature of samples. ZF-$ \mu $SR data reveals a difference in the asymmetry spectra as temperature goes below T$ _{c} $. However, fitting parameters show a gradual increase, ruling out the presence of any spontaneous field below T$ _{c} $, which can otherwise give a sudden increase in relaxation rate at T$ _{c} $. This behavior can be attributed to electronic fluctuations measurable within the $ \mu $SR time scale. It is also noteworthy that, LaNiSi with low ASOC has shown a stronger electronic fluctuation. A recent report on noncentrosymmetric Re$ _{5.5} $Ta by Arushi $ et$ $al. $ (manuscript submitted) has shown spin fluctuation behavior, while similar other Re-based compounds from the family have shown TRSB. These evidences urge for more investigations to elucidate the correlation between ASOC, spin fluctuation, and presence/absence of time reversal symmetry.

\section{Acknowledgments}
R.~P.~S.\ acknowledges Science and Engineering Research Board, Government of India for the Ramanujan Fellowship through Grant No. SR/S2/RJN-832012. We thank Newton Bhabha funding and ISIS, STFC, UK, for the muon beam time to conduct the $\mu$SR experiments [DOI: 10.5286/ISIS.E.67770571].

\end{document}